\documentstyle[12pt]{article}

\def\hybrid{\topmargin -20pt  \oddsidemargin 0pt
      \headheight 0pt   \headsep 0pt
      \textwidth 6.25in 
      \textheight 9.5in 
      \marginparwidth .875in
      \parskip 5pt plus 1pt   \jot = 1.5ex}

\hybrid
\begin{document}
\def\beq{\begin{equation}}
\def\eeq{\end{equation}}
\def\beqa{\begin{eqnarray}}
\def\eeqa{\end{eqnarray}}
\def\beq{\begin{equation}}
\def\eeq{\end{equation}}
\def\beqa{\begin{eqnarray}}
\def\eeqa{\end{eqnarray}}
\sloppy
\newcommand{\be}{\begin{equation}}
\newcommand{\eq}{\end{equation}}
\begin{titlepage}
\begin{center}
\hfill HUB-EP-96/61\\
\vskip .3in
{\bf CLASSICAL ENTROPY OF\ $N=2$ BLACK HOLES:\break 
$The$ $minimal$ $coupling$ $case$}
\vskip .2in
{W. A. Sabra}\footnote{ 
sabra@qft2.physik.hu-berlin.de} \hfill
\vskip 1.2cm
{\em Humboldt-Universit\"at zu Berlin,
Institut f\"ur Physik, 
D-10115 Berlin, Germany}\\
\end{center}
\vskip .2in
\begin{center} {\bf ABSTRACT } 
\end{center}
\begin{quotation}\noindent
We discuss the entropy and the transformation properties of classical
extremal $N=2$ black hole
solutions in supergravity theories associated with 
the minimal coupling models $CP(n-1,1)$.  The entropy is given by a 
manifestly invariant quantity under the embedding of the duality group 
$SU(1,n)$ into $Sp(2n+2)$ which is a symmetry of the classical
BPS mass formula.
\end{quotation}
\end{titlepage}
\vfill
\eject
\newpage
Recently there has been considerable progress in the understanding of 
extremal dyonic black 
holes in $N=2$ supergravity theories using the symplectic formualtion of the 
underlying special 
geometry. In a theory with $N$ vector multiplets, these black holes are 
characterized by a set 
of electric and
magnetic charges $(M_I, N^I)$ $(I=0,\cdots, N)$ where the gauge group of the 
theory is 
given by $U(1)^{N+1}$ and the extra $U(1)$ factor is due to the graviphoton. 
In $N=2$ supersymmetric theories the mass of the BPS states is given by 
\cite{CDFP}
\be
M^2_{BPS}=\vert Z\vert^2.
\eq
where $Z$ is the central charge of the $N=2$ supersymmetry algebra. 
The black hole ADM 
mass is given by the central charge taken at spatial 
infinity 
\be
M^2_{ADM}=\vert Z_\infty\vert^2=\vert Z(z_\infty^A, \bar z_\infty^A)\vert^2.
\eq
Here $z^A(\infty)=z^A_\infty$ $(A=1,\cdots,N),$ are the values of the 
moduli at spatial infinity.
The black hole metric is asymptotically flat at infinity and near the 
horizon the metric 
describes a Bertotti-Robinson universe. In both limits,
the $N=2$ supersmmetry is preserved and the interpolating fields 
configuration between these two maximally 
supersymmetric field configurations breaks half of the 
supersymmetry, which means that one is dealing with BPS states.

It was shown in \cite{fs} that the moduli take fixed points values at 
the horizon 
which depend
only on the electric and magnetic charges and independent of the initial 
values of the moduli
fields. In this case, the Bekenstein-Hawking entropy is given by 
\be
{\cal S}_{BH}(M^I, N_I)={1\over4}A(M^I,N_I)=\pi M^2_{BR}(M^I,N_I)=
\pi\vert Z_{hor}\vert^2.
\eq

Soon after a general principle was discovered, the so-called principle of 
minimal charge, 
which basically states that the 
values of the moduli at the horizon extremize the central charge 
\cite{FerKal1}.
The square of the central charge at the horizon
gives the area of the horizon in terms of the quantum numbers and therefore 
the BH entropy. 
This result comes about as a consequence of unbroken supersymmetry of the 
black hole near 
horizon. Particularly simple black hole solutions are the so-called 
double-extreme dyonic black 
holes, 
those where the moduli fields are constants 
everywhere \cite{KalShmWon} in which case
$M^2_{BR}=M^2_{ADM}$ and the metric is that of Reissner-Nordstr\"om.

In the original analysis of Kallosh, Ferrara and Strominger 
\cite{fs} the 
holomorphic function essential for the superconformal tensor analysis of 
special 
geometry was used. However, the principle of minimal charge uses a 
coordinate free analysis of 
special geometry and therefore independent of the existence of the 
holomorphic function.

In this letter we will study $N=2$ supergravity models with the minimal 
coupling
moduli spaces $SU(1,n)\over U(1)\otimes SU(n)$ and determine the values of 
the moduli fields
at the horizon and the entropies of the corresponding black holes. 
The analysis of the other 
set of 
special K\"ahler manifolds, namely, 
${SU(1,1)\over U(1)}\otimes {SO(2,n)\over SO(2)\times SO(n)}$
is given in \cite{KalShmWon,BKRSW, CLM}. 
Also the quantum corrections for the entropy in these
models has recently been discussed in \cite{BCWKLM, R}.

We first briefly say few words on the symplectic formalism
of $N=2$ supergravity coupled
to $n_V$ vectormultiplets \cite{CDFP}.
The action of $N=2$ supergravity coupled to
$n_V$ vectormultiplets can be determined in terms of the symplectic sections 
$\Omega^t=\pmatrix{X^\Lambda&F_\Lambda}$ $\Lambda=0,\ldots,n_V$ which 
encodes the information about the moduli space as well as the whole 
lagrangian. 
In cases where an $F$ function exists, $F_\Lambda$ is simply the derivative 
of $F$ 
with respect to $X^\Lambda$. Clearly the moduli space is parametrized by the
values of the scalar fields.
The combined set of field equations and Bianchi identities is
invariant under symplectic transformations $\Omega \in Sp(2(n_V+1))$,
which act on the section $\Omega$ as
\be
\left( \begin{array}{c}
X^I \\ F_I \\
\end{array} \right) 
\rightarrow
\Omega 
\left( \begin{array}{c}
X^I \\ F_I \\
\end{array} \right) = 
\left( \begin{array}{cc}
X & Y \\ Z & T \\
\end{array} \right)
\left( \begin{array}{c}
X^I \\ F_I \\
\end{array} \right) \;\;.
\label{st}
\eq
These symplectic transformations are continious at the classical level but 
broken to a discrete
subgroup at the quantum level due to instantons effects.
Letting the $X^I$ be proportional to holomorphic sections $Z^I(z)$ of a
projective $(n+1)$-dimensional space, where $z$ is a set of $n$
complex coordinates, then the $z$ coordinates parametrize a K\"ahler space
with metric $g_{\alpha\bar\beta}=\partial_\alpha\partial_{\bar\beta}
K$, where $K$, the K\"ahler potential is expressed by
\beqa
K&=&-\log i\Big[Z^I\bar F_I(\bar Z)-F_I(Z)\bar Z^I\Big]\cr
&=&-\log i\pmatrix{Z^I&F_I(Z)}\pmatrix{\bf0&\bf1\cr -\bf1&\bf0}
\pmatrix {\bar Z^I\cr \bar F_I(\bar Z)},\cr
X^I&=&e^{K/2}Z^I,
\bar X^I=e^{K/2}\bar Z^I.
\eeqa
The so-called special coordinates correspond to the choice
\be
z^\alpha={X^\alpha\over X^0}; \qquad Z^0(z)=1,
\quad Z^\alpha(z)=z^\alpha.
\eq
The central charge $Z$ depends on the quantum numbers, electric and magnetic 
charges, 
as well as the moduli and is given by
\be 
M_{BPS}^2 =|Z|^2 = |M_I X^I + N^I F_I|^2.
\eq
>From the knowledge of the embedding of the duality group of the moduli space 
into 
the symplectic group
$Sp(2n_V+2)$ \cite{sabra}, the relation between $F_I$ and 
$X^I$ can be determined. Whether an $F$ function exists or not 
depends very much on the choice of the embedding. In this formulation, the 
embedding 
contains the full information about the lagrangian of the theory \cite{GZ}.

We now specialize to the $CP_{n-1,1}\equiv {SU(1,n)\over SU(n)\times U(1)}$ 
models and discuss 
their special geometry
and symplectic sections.  
The isometry group of these cosets is given by the group
$SU(1,n).$ If we represent an element of $SU(1, n)$ by an $(n+1)\times (n+1)$
complex
matrix $M$ satisfying
\be
{M}^\dagger\eta {M}=\eta,\quad \det M={1}, 
\label{mars}
\eq with
$\eta$ the
constant diagonal metric with signature $(+, -,\cdots, -)$,
and decompose the matrix $M$ into its real and imaginary part,
\be
{M}={U}+i{V},
\label{dec}
\eq
then (\ref{mars}) implies
for the real $(n+1)\times (n+1)$ matrices $U$ and $V,$ the following
relations
\beq
{U}^t\eta {U}+{V}^t\eta {V}=\eta,\qquad 
{U}^t\eta {V}-{V}^t\eta {U}=0.
\label{cybill}
\eq

Recall that an element $\Omega$ of $Sp(2n+2)$ is a
$(2n+2)\times (2n+2)$ real matrix
satisfying
\be
{\Omega}^tL{\Omega}= L,\qquad
L=\pmatrix{{0}&{1}\cr -{1}&{0}},
\label{pinhead}
\eq
If we write 
\be
\Omega=\pmatrix{{A}&{B}\cr {C}&{ D}}
\label{swansha}
\eq
where the matrices $A$, $ B$ $C$ and $D$ are
$(n+1)\times (n+1)$ matrices, then in  terms of these block matrices,
(\ref{pinhead}) implies the following conditions
\be
{A}^t{C}-{C}^t{A}={0},\quad
{A}^t{D}-{C}^t{B}={1},\quad
{ B}^t{D}-{D}^t{B}={0}.\label{ero}
\eq 
An embedding $\Omega_e$ of $SU(1,n,{\hbox{\bf Z}})$
into the symplectic group $Sp(2n+2)$ is given in components by
\be
{A}={U}, \quad {C}=-\eta {V}, \quad {B}={V}\eta,
\quad {D}=\eta {U}\eta.\label{com}
\eq

Introduce the symplectic
section $\pmatrix{X^\Lambda&F_\Lambda}$ which
transforms as a vector under the symplectic transformations
induced by ${\Omega}_e$. These transformation rules can then be used
to determine the relation between $F_\Lambda$ and the
coordinates $X^\Lambda$.
In components, these transformations are given by
\be
X \rightarrow {U} X+{V}\eta\partial F,\qquad
\partial F\rightarrow -\eta {V} X+\eta {U}\eta \partial F,
\label{tra}
\eq
where $X$ and $\partial F$ are $(n+1)$-dimensional vectors
with components $X^\Lambda$ and $F_\Lambda$ respectively.
It can be seen from the relations (\ref{tra}) that $\partial F$
can be identified with
$i\eta X$, and as such, a holomorphic prepotential $F$ exists
and is given,
in terms of the
coordinates $X$, by
\be
F={i\over2}X^t\eta X.
\label{Newyork}
\eq
In this case, the complex vector $X$ transforms as
\be
X\rightarrow ({U}+i{V})X={M}X,\label{dd}\eq
which implies that $X$ is proportional to the complex coordinates
parametrizing the ${SU(1, n)\over U(1)\times SU(n)}$ coset, and satisfying the following 
relation,
\be
\phi^\dagger\eta\phi={1}, \qquad\hbox{where}\quad
\phi=\left(\matrix{\phi^0\cr\vdots \cr\phi^{n+1}}\right),\label{nem}
\eq
and are parametrized in terms of
unconstrained coordinates $z^\alpha$ by 
\be
\phi^0={1\over \sqrt Y},\quad
\phi^j={z^\alpha\over \sqrt Y}, \quad \alpha=1, \cdots, n,\label{berlin}
\eq
where $Y={1-\sum_\alpha z^\alpha\bar z^\alpha}.$
Here we identify $X$ with the complex vector ${1\over\sqrt2}\phi$.
The special coordinates in this case are given by
$z^\alpha$ and the K\"ahler potential is
given by
\be
K=-log(2-2\sum_\alpha z^\alpha{\bar z}^\alpha).\label{so}
\eq

A different embedding of $SU(1,n)$ into $Sp(2n+2)$ of course leads to
a different relation between $F_\Lambda$ and $X^\Lambda.$
In fact once an embedding $\Omega_e$ is specified,
then for all elements $S\in Sp(2n+2),$
the matrix
\be
\Omega'_e=S\Omega_e S^{-1},\label{ae}
\eq
provides another embedding
with a corresponding symplectic section. As an example, consider the element
\be
S_1=\pmatrix{{\Sigma}&{0}\cr{0}&{\Sigma}},\quad
{\hbox{with}}\qquad
{\Sigma}=\pmatrix{{1\over\sqrt2}{\sigma}&{0}\cr{0}&{1}},
\qquad
\sigma=\pmatrix{1&1\cr 1&-1}.\label{sis}
\eq

Using (\ref{ae}) and (\ref{sis}), another
embedding of $SU(1,n)$ into $Sp(2n+2)$ can be obtained and is
given by
\be
\Omega_e'=\pmatrix{{\Sigma}{U}{\Sigma}&{\Sigma}
{V}\eta{\Sigma}\cr -{\Sigma}\eta{V}{\Sigma}&
{\Sigma}\eta{U}\eta{\Sigma}}.\label{nst}
\eq
The new section $\pmatrix{X'&\partial F'}$ is related to 
$\pmatrix{X&\partial F}$ as follows
\beqa
X'^0={1\over\sqrt2}(X^0+X^1),\quad
X'^1={1\over\sqrt2}(X^0-X^1),\quad
X'^j=X^j,\qquad j=2, \cdots, n\cr
F'_0={i\over\sqrt2}(X^0-X^1)=iX'^1,\quad
F'_1={i\over\sqrt2}(X^0+X^1)=iX'^0,\quad
F'_j=-iX^j=-iX'^j.\label{yael}
\eeqa
>From (\ref{yael}), it can be easily seen that there exists a
holomorphic prepotential
$F'$ which can be expressed in terms of $X'$ by
\be
F'={i}\Big(X'^0X'^1-{1\over2}\sum_{j=2}^n (X'^j)^2\Big).\label{nf}
\eq
For this parametrization, we have
\be
Z'^0=1,\quad Z'^1={1-z^1\over 1+z^1}, \qquad Z'^j={\sqrt 2 z^j\over
1+z^1},\label{sue}
\eq
and the K\"ahler potential is given by
\be
K=-\log (Z'^1+\bar Z'^1-\sum_{j}Z'^j\bar Z'^j).\label{lon}
\eq
We now consider the simplest case, 
namely $N=2$ supergravity with one scalar field (the modulus) parametrizing 
the coset $SU(1,1)\over U(1)$ \cite{KalShmWon}. 
For the embedding as defined in (\ref{com}), $X$ transform under duality as 
an $SU(1,1)$ vector.
\be
\pmatrix{X^0\cr X^1}\rightarrow \pmatrix{z_1&\bar z_2\cr z_2&\bar z_1}
\pmatrix{X^0\cr X^1},\quad \vert z_1\vert^2-\vert z_2\vert^2=1.\label{demons}
\eq
If we represent the modulus by the special coordinate $t={X^1\over X^0}$, 
then under duality
\be
t\rightarrow {z_2+\bar z_1t\over z_1+\bar z_2t}.
\label{pain}
\eq
The K\"ahler potential is given by
\be
K= -\log(2-2t\bar t),\label{napalm}
\eq
and the  
BPS mass formula in terms of $t$ is given 
by
\be
M_{BPS}^2 =|Z|^2 =e^K |M_0+M_1t +i N^0-iN^1t|^2 = e^K|{\cal M}|^2,
\eq
where $\cal M$ is the so-called holomorphic mass,
$Z$
is the central charge of the $N=2$ 
supersymmetry algebra and  
$M_0, M_1$ and $N^0, N^1$ are the symplectic
quantum numbers related to electric and magnetic charges of the $U(1)\times U(1)$ 
gauge group.

Define 
\be
m_c=M_0+iN^0, \, n_c=iN^1-M_1,\label{before}
\eq
in terms of which, the central charge formula can be rewritten in a simpler form
\be
M_{BPS}^2 =|Z|^2 ={|m_c-n_ct|^2\over 2(1-t\bar t)}.
\label{central}
\eq

The moduli fields near the horizon are driven to fixed values determined via the relation
\be
D_iZ=(\partial_i+{1\over2}K_i)Z=0\Longleftrightarrow\quad \partial_i\vert Z\vert=0.
\eq
Substituting the extremal values of the moduli into the square of the 
central charge one obtains the Bekenstein-Hawking entropy.

The extremization of the central charge as given in (\ref{central}) gives 
\be
\bar t={n_c\over m_c},
\eq
thus the entropy is given by
\be
{\cal S}={\pi\over2}\Big({\vert m_c\vert^2-\vert n_c\vert^2}\Big)={\pi\over2}
\pmatrix{\bar m_c&\bar n_c}
\pmatrix{1&0\cr 0&-1}\pmatrix{m_c\cr n_c}.
\label{tea}
\eq
The entropy is invariant under the duality transformations, this can be 
easily verified 
by noticing that it can be rewritten as 
\be
{\cal S}={\pi\over2}\pmatrix{M&N}\pmatrix{\eta&0\cr 0&\eta}\pmatrix{M\cr N}.
\label{horror}
\eq
Under the embedding of the duality group in $Sp(4)$,  
the quantum numbers transform as
\be
\pmatrix{M\cr N}\rightarrow \pmatrix {\eta U\eta&\eta V\cr -V\eta&U}\pmatrix{M\cr N}.
\eq
Using the above transformation and the conditions (\ref{cybill}) satisfied by $U$ and $V$, 
it can be seen that the entropy is duality invariant.
Moreover, the duality invariance is also manifest in the form (\ref{tea}) after noticing that 
$\pmatrix{m_c\cr n_c}$ transforms as a vector under the $SU(1,1)$ duality 
transformation.

Let us now analyse the same model in the paramerization $({X'}^0,{X'}^1)$ of the moduli space
corresponding to the embedding
(\ref{nst}) for the $SU(1,1)$ case. Represent the modulus by a new special coordinate
$\textstyle T={\textstyle {X'}^1\over\textstyle {X'}^0}.$
If we write
\be
(z_1+z_2)=(d+ic),\quad (z_1-z_2)=(a-ib),
\label{scatter}
\eq
then the condition $z_1\bar z_1-z_2\bar z_2=1,$
implies that $ad-bc=1.$
Using (\ref{nst}) we obtain
the following embedding in $Sp(4),$
\be
\Omega'_{SU(1,1)}
=\pmatrix{d&0&c&0\cr 0&a&0&-b\cr b&0&a&0\cr 0&-c&0&d}.\label{li}
\eq
This gives using (\ref{nf}), the familiar $SL(2,{\hbox{\bf Z}})$ transformation
for the modulus $T$ 
\be
T\rightarrow {aT-ib\over icT+d}.\label{red}
\eq
In this parametrization, we have
\be
M_{BPS}^2 =|Z|^2 ={1\over{T+\bar T}} |m_0+m_1T +i n^0T+in^1|^2,
\eq
Define 
\be
\alpha_c=m_0+in^1, \, \beta_c=m_1+in^0,
\eq
in terms of which, the above equation can be rewritten in the form
\be
M_{BPS}^2 =|Z|^2 ={1\over{T+\bar T}} |\alpha_c+\beta_cT|^2,
\eq
Upon extremization we get
\be
\bar T={\alpha_c\over\beta_c},
\eq
and the entropy is thus given by
\be
{\cal S}=\pi(\alpha_c\bar\beta_c+\bar\alpha_c\beta)=\pi\pmatrix{\bar\alpha_c&\bar\beta_c}
\pmatrix{0&1\cr 1&0}\pmatrix{\alpha_c\cr\beta_c}.
\eq
Obviously, this result can also be obtained from (\ref{horror}) by performing a symplectic 
transformation, which connects the
two parametrization,
on the quantum numbers. This gives
\be
\pmatrix{M&N}\pmatrix{\eta&0\cr0&\eta}\pmatrix{M\cr N}=
\pmatrix{m&n}\pmatrix{\Sigma\eta\Sigma&0
\cr 0&\Sigma\eta\Sigma}\pmatrix{m\cr n}=\pmatrix{\bar\alpha_c&\bar\beta_c}
\pmatrix{0&1\cr 1&0}\pmatrix{\alpha_c\cr\beta_c}.
\eq

Finally, we analyze the case in which an $F$ function does not exist 
\cite{CDFP,P}. In this 
case the corresponding symplectic section $\pmatrix{X''&\partial F''}$ is 
related to 
$\pmatrix{X&\partial F}$ by
\be
\pmatrix{X''\cr \partial F''}=\pmatrix{A&B\cr -B&A}
\pmatrix{X\cr \partial F}
\quad
\hbox {where}\quad A={1\over\sqrt2}\pmatrix{1&1\cr 0&0},\quad 
B={1\over\sqrt2}\pmatrix{0&0\cr -
1&1}
\eq

In this parametrization, the new set of electric and magnetic quantum 
numbers,$(a,b),$ are given by
\be
M^t=a^tA-b^tB,\quad N^t=a^tB+b^tA.
\eq
The entropy is given by
\be
{\cal S}=\pi(a_0b^1-a_1b^0)={\pi\over2}
\pmatrix{a_0&a_1&b^0&b^1}\pmatrix{0&0&0&1\cr 0&0&-1&0\cr 0&-1&0&0\cr 1&0&0&0}
\pmatrix{a_0\cr a_1\cr b^0\cr b^1}.
\eq

We now consider the most general case, $SU(1,n)\over SU(n)\times U(1)$, 
and analyze the 
system using the embedding defined in (\ref{com}). The holomorphic 
function in this case 
is given by
\be
F={i\over2}\Big((X^0)^2-(X^1)^2-\sum_{i=2}^{n-1}(X^i)^2\Big)
.\eq 
The K\"ahler potential is given by 
\be K=-\log2(1-t\bar t-\sum_i{\cal A}^i\bar{\cal A}^i)
\eq
The BPS mass formula (central charge) can be written in the form 
\be
M_{BPS}^2={\vert m_c-n_ct-Q_{ic}{\cal A}^i\vert^2\over
2(1-t\bar t-\sum_i{\cal A}^i\bar{\cal A}^i)}
\eq
where $m_c$, $n_c$ are as defined in (\ref{before}), $Q_{ic}$ and the 
moduli are defined by
\be
Q_{ic}=iN^i-M_i,\quad t={X^1\over X^0},\quad {\cal A}^i={X^i\over X^0}.
\eq

The extremization of the central charge gives the following equations for 
the moduli fields in terms of the quantum numbers
\be 
{\bar t}={n_c(1-{\cal A}^i\bar{\cal A}^i)\over m_c-Q_{ic}{\cal A}^i},\qquad
{\bar {\cal A}}^i={Q_{ic}(1-t\bar t)\over m_c-n_c t}
\eq
whose solution is given by
\be
{\bar t}={n_c\over m_c},\qquad {\bar{\cal A}}^i={Q_{ic}\over m_c}
\eq
and the entropy is thus given by
\be
{\cal S}={\pi\over2}\Big(\vert m_c\vert^2-\vert n_c\vert^2-
\vert Q_{ic}\vert^2\Big)=
{\pi\over2}\pmatrix{M&N}\pmatrix{\eta&0\cr 0&\eta}\pmatrix{M\cr N}.
\eq
The above formula is universal for all the $CP(n-1,1)$ cosets with the 
appropriate 
dimensionality of the vector defining the magnetic and electric charges and 
the metric $\eta$.
In order to get the entropy in other parametrization of the moduli space, 
one need only 
to know the relation between the set of charges. Suppose that we have a 
parametrization 
connected to the one we discussed by a symplectic transformation $S$, and 
with the 
quantum numbers $\pmatrix{M'&N'}$, then it can be seen that 
$\pmatrix{M&N}=\pmatrix{M'&N'}S$
and the entropy in terms of  $\pmatrix{M'&N'}$ is thus given by
\be
{\cal S}=
{\pi\over2}\pmatrix{M'&N'}S\pmatrix{\eta&0\cr 0&\eta}S^t\pmatrix{M'\cr N'}.
\eq

{\bf Acknowledgement}\
This work was partly supported by PPARC. I would like to thank 
D. L\"ust for useful discussions.

\end{document}